\documentstyle[preprint,aps]{revtex}
\begin{document}
\input epsf
\title{Invaded cluster algorithm for equilibrium critical
points}
\author{J. Machta, Y. S. Choi, A. Lucke and T. Schweizer}
\address{Department of Physics and
Astronomy, University of Massachusetts,  Amherst, MA 01003-3720}
\author{L. V. Chayes}
\address{Department of Mathematics, University of California, Los
Angeles, CA 90095-1555 }

\maketitle
\begin{abstract}
A new cluster algorithm based on invasion percolation is described.
The algorithm samples the critical point of a spin system without
a priori knowledge of the critical temperature and provides an
efficient way to determine the critical temperature and other
observables in the critical region.  The method is illustrated for
the two- and three-dimensional Ising models.  The algorithm
equilibrates spin configurations much faster than the closely
related Swendsen-Wang algorithm.

\end{abstract}
\narrowtext
Enormous improvements in simulating systems near critical points
have been achieved by using cluster algorithms~\cite{SwWa,Wolff}.
In the present paper we describe a new cluster method which has the
additional property of `self-organized criticality.'  In particular,
the method can be used to sample the critical region of various spin
models without the need to fine tune any parameters (or know them in
advance).  Here, as in other cluster algorithms, bond clusters play a
pivotal role in a Markov process where successive spin
configurations are generated using the Fortuin-Kasteleyn~\cite{FoKa}
representation to identify clusters of spins for flipping.  However,
the clusters themselves are identified using {\it invasion
percolation}. The new algorithm is closely related to the
Swendsen-Wang (SW)  algorithm~\cite{SwWa} and
may be adapted for a wide range of systems. For purposes of
illustration, in this work we will consider the Ising
model.

Let us first recall the SW algorithm as applied to an Ising system.
Starting from a spin configuration, satisfied bonds--those
connecting spins that are of the same type--are occupied with
probability, $p(\beta)=1-e^{-\beta}$ where $\beta=J/k_B T$ is the
coupling strength. Unsatisfied bonds are never occupied.  Clusters
of sites connected by occupied bonds are locked into the same
spin-type and all clusters (including isolated sites) are
independently flipped with probability $1/2$.  The SW algorithm
samples the canonical ensemble of the spin system at coupling
$\beta$ and/or the random cluster (bond configuration) ensemble with
parameter $p$.  At $T_c$ the SW algorithm is far more efficient
than single spin-flip methods because the flipped clusters are also
critical droplets~\cite{CoKl}\/.

Here we propose using invasion
percolation~\cite{Ham,Vi,WiWi,ChKo,WiBa,ChChNe} to generate the bond
clusters for the spin flips.  In the usual invasion  percolation,
random numbers are independently assigned to the bonds of the
lattice.  Growth starts from one or more seed sites and at each step
the clusters grow by the addition of the perimeter bond with the
smallest random number.  If a single cluster grows indefinitely on
an infinite lattice,  its large scale behavior is presumed to be that
of the ``incipient infinite cluster'' of ordinary percolation.  In
particular, the fraction of perimeter bonds accepted into the
growing cluster approaches the percolation threshold,
$p_c$~\cite{WiBa,ChChNe}. Invasion percolation is thus a
self-organized critical phenomenon.

For the present, we modify invasion percolation in two ways.  First,
we initiate cluster growth at {\em all}\/ lattice sites.  Consider
this change for ordinary invasion percolation: every bond is
initially a perimeter bond and the invasion process consists of
collecting bonds in a given random order.  Initially every site is a
cluster and in most steps of the growth process two smaller clusters
are combined into a single larger cluster.  The growth process is
terminated when some cluster first spans the system. Let $f$ be the
fraction of bonds accepted during the growth process and $L$ the
system size. As $L \rightarrow \infty$, $f$ approaches $p_c$, the
ordinary percolation threshold for the corresponding lattice.  An
intuitive argument for this (that can be made precise and in fact
goes back to Ref.\ \cite{Ham}) is as follows: if $f$ were smaller
than $p_c$ the probability of a spanning cluster would vanishes as
$L\rightarrow \infty$ while if  $f>p_c$, with probability
approaching one, many disjoint subsets of the {\em same} cluster
achieve spanning.  Thus, for large systems $f$ is close to $p_c$.

The second modification, which is the cornerstone of this Letter,
correlates invasion percolation to an underlying spin
configuration.  As in the SW algorithm, this is done by allowing
cluster growth along only satisfied bonds.

The new method, which we call the invaded cluster (IC) algorithm,
works as follows.  Starting with an Ising spin configuration, $S$,
the bonds of the lattice are given a random order. Correlated
invasion percolation clusters are grown as described above until one
of the  clusters spans the system.  (In the present implementation,
a cluster is counted as spanning when the maximum separation in one
of the $d$ directions for some pair of points in the cluster is the
system size, $L$.)  After the growth process is terminated, each
cluster is flipped with probability 1/2 yielding a new spin
configuration, $S^\prime$.  The bonds are then randomly reordered
and the process begins anew.

\begin{figure}
\centering
\epsfysize=5.5truein
\epsffile{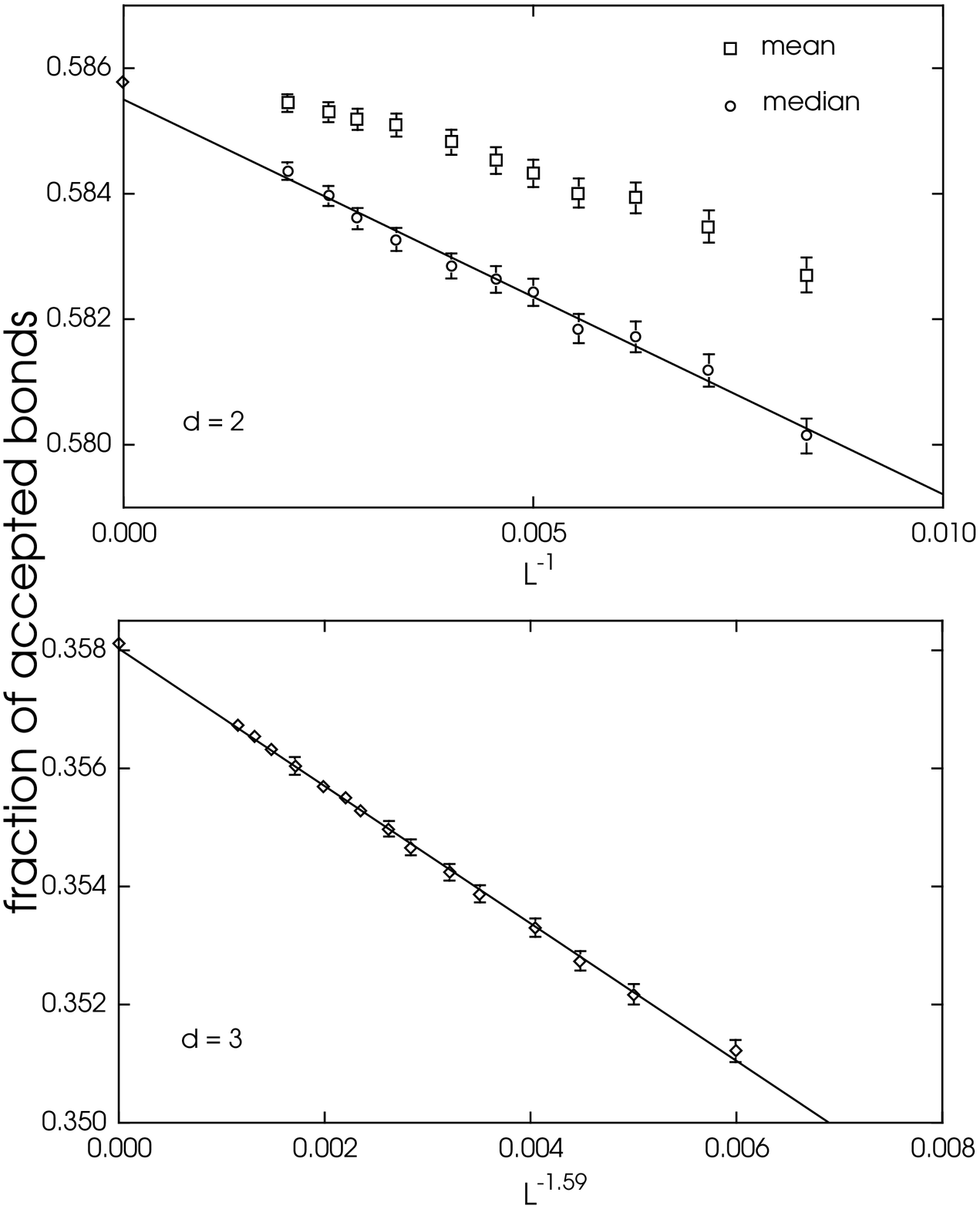}
\caption{\label{2f}  The fraction of accepted bonds for two and
three dimensions vs $L^{-1/\nu}$. For two dimensions the mean
 and median values of $f$ are shown.  The solid lines are linear
fits through the data and the known values of $p(\beta_c)$ are marked
on the vertical axes with diamonds.}
\end{figure}

We tested the algorithm on the nearest neighbor, ferromagnetic Ising
model on square and simple cubic lattices with periodic boundary
conditions.  The computation time per Monte Carlo step scales
linearly in the number of spins but with a somewhat larger prefactor
than for the SW algorithm.  After equilibration for 200 steps
starting from the spin up state, statistics were collected on the
energy, $E$ and the ratio, $f$ of the number of accepted bonds to
the number of satisfied bonds. For each system size we collected
statistics for the order of $10^4$ Monte Carlo steps.  We first
discuss the results for two-dimensional systems with sizes up to
$500^2$.  In Fig.\ \ref{2f} we show the mean and median~\cite{tail}
values of $f$ plotted against $1/L$.  A linear fit through the
median data extrapolates to 0.5855 compared with the exact value,
$p(\beta_c)=1-e^{-\beta_c}= 0.58579\ldots$\/.

\begin{figure}
\centering
\epsfysize=6.truein
\epsffile{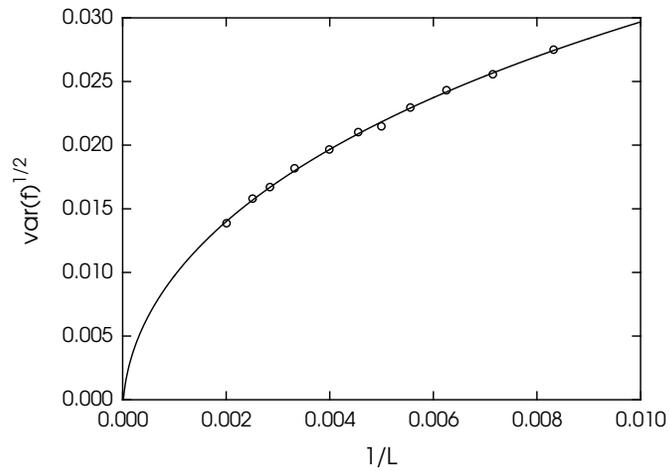}
\caption{\label{2varf} The standard deviation of $f$ vs $1/L$ for
the two-dimensional Ising model.  The solid line is the least
squares fit to the form $c_0 + c_1 L^{-1/2}+c_2 L^{-1}$.  }
\end{figure}

\begin{figure}
\centering
\epsfysize=5.5truein
\epsffile{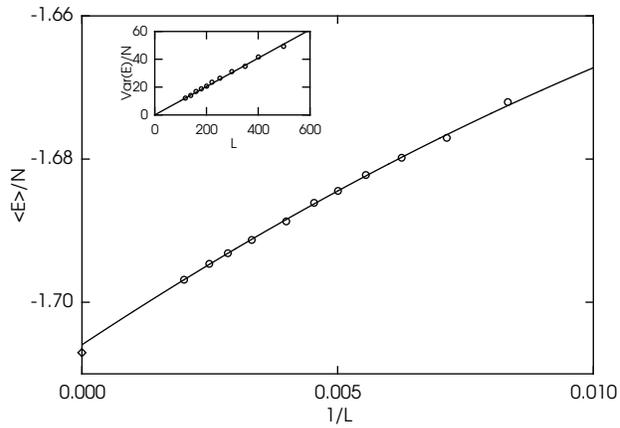}
\caption{\label{2energy} $\langle E \rangle/N$ vs $1/L$ for the
two-dimensional Ising model.  The solid line is a fit to the form
$e_0 + e_1 L^{-1}+e_2 L^{-2}$ and the exact infinite volume result
is indicated by a diamond on the vertical axis. The inset shows
var$(E)/N$ vs $L$.  The solid line is a linear fit through the
data.}
\end{figure}

In Fig.\ \ref{2varf}, we plot the standard deviation of $f$ as a
function of $1/L$.  The solid line is a fit to a function of the
form $c_0 + c_1 L^{-1/2}+c_2 L^{-1}$ which yields $c_0=-0.0014$.
Figures\ \ref{2f}  and \ref{2varf} are thus consistent with the
hypothesis that the distribution of $f$ approaches a delta function
at $p(\beta_c)$ as $L\rightarrow \infty$ with a scaling that is
given, approximately, by  $L^{-1/2}$.  The average energy per spin,
$\langle E \rangle/N$ is shown in Fig.\ \ref{2energy} and plotted
against the inverse of the system size, $1/L$.  The best fit to the
form $e_0 + e_1 L^{-1}+e_2 L^{-2}$ yields $e_0=-1.706$ in comparison
to the exact result, $-1.7071$. The variance of the total energy
divided by the number of spins is shown in the inset of Fig.\
\ref{2energy}.  In the canonical ensemble, var$(E)/N$ is
proportional to the specific heat and diverges logarithmically in
$L$ whereas here we find that this quantity diverges linearly in
$L$.  It is clear that the IC algorithm does not sample the
canonical ensemble.

The results for the three-dimensional Ising model up to system size
$70^3$ are qualitatively similar to the two-dimensional results
except that the finite size scaling behavior for $f$ is controlled by
the three-dimensional Ising correlation length exponent, $\nu\simeq
.63$.  In Fig.\ \ref{2f} we plot the mean value of $f$ against
$L^{-1/\nu}$.  The solid line in the figure is the least squares
linear fit to all the data which yields an intercept at .35803  in
comparison to the accepted value~\cite{FeLa},
$p(\beta_c)\simeq.35810$.  The standard deviation of $f$ vanishes
with a leading behavior that is well fitted to  $L^{-1/2\nu}$.  The
mean energy per spin  extrapolates to  $-2.0$.

Why does the IC algorithm work?  We do not have a rigorous
proof that as the system size goes to infinity the distribution for
$f$ peaks at $p(\beta_c)$ or that observables such as the energy
density converge to their limiting (infinite volume) values at
criticality.  Nonetheless we can give some heuristic arguments
supporting the validity of the algorithm.

The discussion is based on the observation that each iteration of
the invaded cluster algorithm is identical to one iteration of the
SW algorithm with $p=f$.  Suppose we start with an infinite (or
huge) spin configuration that is already typical of the critical
point. On the basis of current understanding,
$p(\beta_c)$ corresponds to the threshold for the formation of
large--scale bond clusters in the associated correlated bond
percolation problem on this configuration. Let us now parallel the
arguments given above for uncorrelated invasion percolation. On a
finite system, in a critical spin configuration, the fraction $f$ of
accepted satisfied bonds will be close to $p(\beta_c)$ when some
cluster first  spans the system.  Hence, when the spin clusters get
flipped, the new configuration should still be typical of
criticality. It thus follows that if the invaded cluster algorithm
starts from a  critical spin configuration, it behaves like the SW
algorithm with a temperature that fluctuates near $T_c$ and
therefore remains near criticality.

It is also clear that the IC algorithm moves the spin configuration
toward criticality if it is started in either the high or low
temperature phase.  Suppose the spin configuration is in the low
temperature phase. Here the portion of satisfied bonds is greater
than the critical value and due to this relative abundance, a
smaller fraction is needed to produce a spanning cluster than in the
case of a critical spin configuration. For example, in the extreme
case of the zero temperature configuration, one obtains  $f \simeq
p_c$, the ordinary percolation point  which is of course
significantly smaller than $p(\beta_c)$. Writing $f=1-e^{-\beta}$,
this corresponds to an iteration of the SW algorithm at $T>T_c$ and
the system is pushed toward higher temperature. Conversely, if the
spin system is in the high temperature phase, there are not enough
satisfied bonds and spanning will occur for $f>p(\beta_c)$,
corresponding to an iteration of the SW algorithm at a temperature
less than the critical temperature.

These arguments suggest that, in finite volume, the stationary
distribution of the IC algorithm is close to (although not identical
to) the canonical ensemble at $\beta_c$ and/or the corresponding
Fortuin-Kasteleyn random cluster distribution at $p(\beta_c)$.  We
will refer to the distribution sampled by the algorithm as the
invaded cluster ensemble. Let us further suppose that the
distribution for $f$ becomes sharp as $L\rightarrow \infty$ and that
the volume fraction  of the spanning cluster tends to zero in this
limit.  It then follows that in the invaded cluster ensemble, the
distribution functions of all local observables, e.g. spin
correlation functions or cluster size distributions, will converge
to their infinite volume critical point distributions.  The critical
exponent, $\tau$($=2+\beta/(\beta+\gamma)$) can be obtained from the
cluster size distribution.  Our measurements of the cluster size
distribution for the two- and three-dimensional Ising models are
consistent with the accepted values of $\tau$.  Another independent
exponent can presumably be extracted via finite-size scaling. On the
other hand, we must emphasize that finite-volume fluctuations in the
invaded cluster ensemble such as var$(E)$ need not have the same
value as in the canonical ensemble and cannot be interpreted as
thermodynamic response functions.

\begin{figure}
\centering
\epsfysize=5.5truein
\epsffile{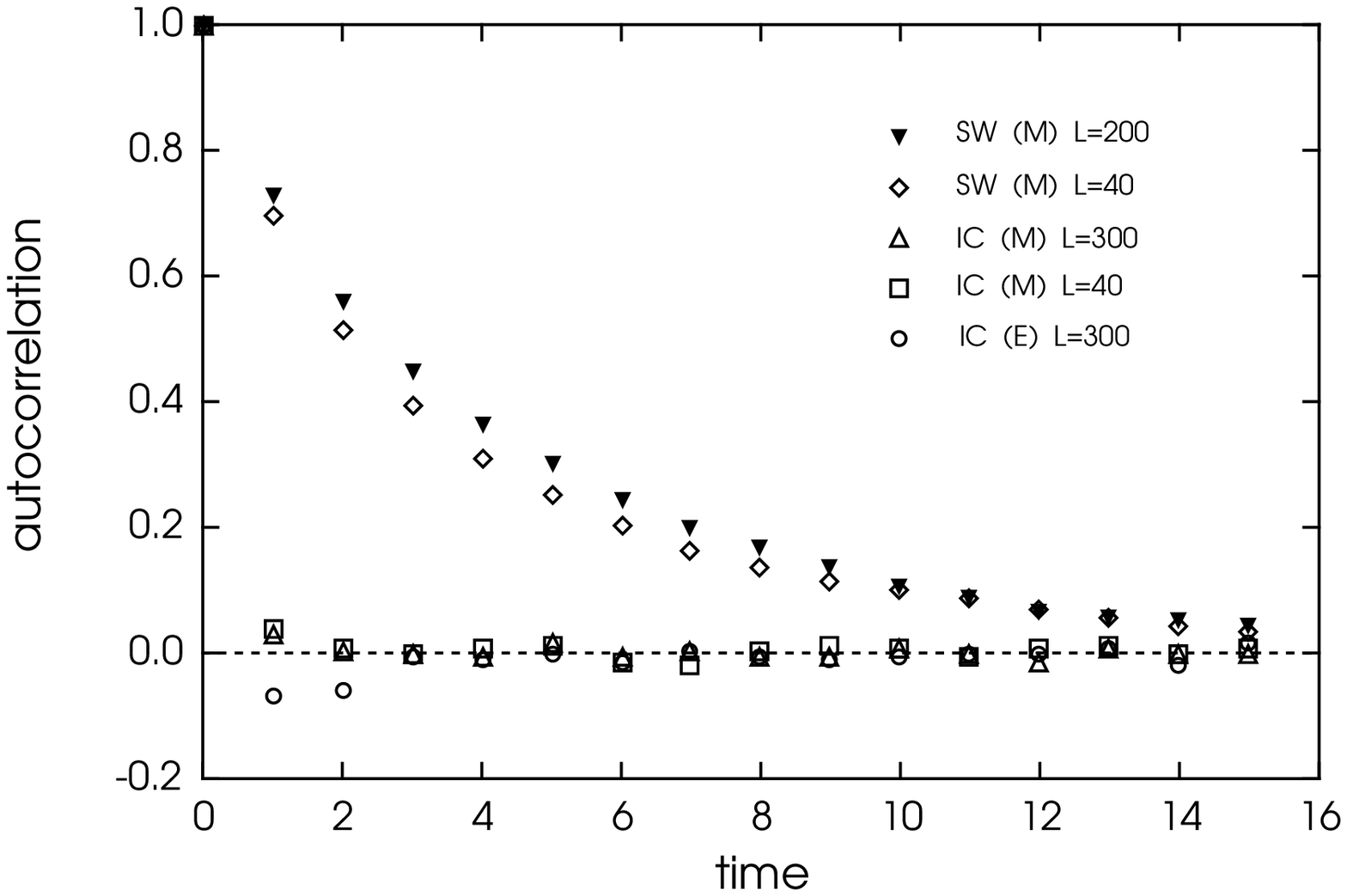}
\caption{\label{auto} The normalized magnetization (M) and energy
(E) autocorrelation functions for Swendsen-Wang (SW) and invaded
cluster (IC) dynamics for the two-dimensional Ising model.}
\end{figure}

We measured the normalized energy and magnetization autocorrelation
functions.  The energy autocorrelation function is defined by
$\langle(E(t)-\langle E\rangle)(E(0)-\langle
E\rangle)\rangle/$var$(E)$ with $t$ the number of iterations
of the algorithm.  Results for two dimensions are plotted in Fig.\
\ref{auto} and compared to the SW algorithm. The energy and
magnetization are almost fully decorrelated in a single Monte Carlo
step!  Results for the three-dimensional Ising model are similar.
The negative overshoot in the energy autocorrelation is
consistent with the negative feedback mechanism described above
and the latter suggest why the algorithm is so fast.
Consider again the example of an initial spin configuration at zero
temperature: one iteration of the SW algorithm at $\beta_c$ yields
a bond percolation configuration at $p(\beta_c) > p_c$ which still
maintains a considerable degree of low temperature order. In
particular, the average magnetization per site is still
appreciable.  By contrast, the magnetization after one step of the
IC algorithm will have essentially vanished.  If the same type of
reasoning is applied to more general configurations, the conclusion
is that the IC algorithm drives a system to criticality faster than
the SW algorithm with $p=p(\beta_c)$. It is tempting to speculate
that in some cases invaded cluster dynamics has no critical
slowing down.

The invaded cluster algorithm should find many uses. The extremely
rapid equilibration time suggests it may be
the best approach for high precision simulations of the critical
region of large spin systems.  Using the embedding method of
Ref.~\cite{Wolff}, continuous spin models may be
simulated. The algorithm may also be useful for first
order transitions, preliminary results for the $d=3, q=3$ Potts
models indicates that the transition temperature is correctly
located.  The IC algorithm should also prove useful for quenched random
ferromagnetic systems where the critical temperature--which depends
on the details of the disorder distribution--is often difficult to
pin down.

We acknowledge useful conversations with M. S. Cao during
the early stages of this work. This work was supported by NSF Grants
DMR-93-11580 and DMS-93-02023.


\vspace*{1.in}

\begin{thebibliography}{10}

\bibitem{SwWa}
R.~H. Swendsen and J.-S. Wang.
\newblock Nonuniversal critical dynamics in {Monte} {Carlo} simulations.
\newblock {\em Phys. Rev. Lett.}, 58:86, 1987.

\bibitem{Wolff}
U.~Wolff.
\newblock Collective {M}onte {C}arlo updating for spin systems.
\newblock {\em Phys. Rev. Lett.}, 62:361, 1989.

\bibitem{FoKa}
C.~M. Fortuin and P.~M. Kasteleyn.
\newblock On the random-cluster model.
\newblock {\em Physica}, 57:536, 1972.

\bibitem{CoKl}
A. Coniglio and W. Klein.
\newblock Clusters and {Ising} critical droplets: A
renormalisation group approach.
\newblock {\em J. Phys. A: Math. Gen.}, 13:2775, 1980.

\bibitem{Ham} J.~M. Hammersley.
\newblock A Monte Carlo solution of percolation in the cubic
lattice.
\newblock {\em In:  Methods in computational Physics, Vol. I}.
\newblock Academic Press, New York, 1963.

\bibitem{Vi}
T.~Vicsek.
\newblock {\em Fractal Growth Phenomena}.
\newblock World Scientific, Singapore, 1992.

\bibitem{WiWi}
D.~Wilkinson and J.~F. Willemsen.
\newblock Invasion percolation: A new form of percolation theory.
\newblock {\em J. Phys. A: Math. Gen.}, 16:3365, 1983.

\bibitem{ChKo}
R.~Chandler, J.~Koplick, K.~Lerman, and J.~F. Willemsen.
\newblock Capillary displacement and percolation in porous media.
\newblock {\em Journal of Fluid Mechanics}, 119:249, 1982.

\bibitem{WiBa}
D.~Wilkinson and M.~Barsony.
\newblock Monte {Carlo} study of invasion percolation clusters in two and three
  dimensions.
\newblock {\em J. Phys. A: Math. Gen.}, 17:L129, 1984.

\bibitem{ChChNe}
J.~T. Chayes, L.~Chayes, and C.~M. Newman.
\newblock The stochastic geometry of invasion percolation.
\newblock {\em Comm. Math. Phys.}, 101:383, 1985.

\bibitem{tail} The difference between the mean and median values
results from a tail in the distribution of $f$ extending toward
$f=1$.
For the three-dimensional Ising
model, the mean and median values of $f$ are nearly the same.

\bibitem{FeLa}
A.~M. Ferrenberg and D.~P. Landau.
\newblock Critical behavior of the three-dimensional {I}sing model: A
  high-resolution {M}onte {C}arlo study.
\newblock {\em Phys. Rev. B}, 44:5081, 1991.




\end{thebibliography}
\end{document}